\documentclass[12pt,titlepage]{article}
\usepackage{amssymb}
\usepackage{epsfig}
\usepackage{bm}
\font\grande=cmr10 scaled \magstep4
\font\medio=cmr10 scaled \magstep2
\outer\def\beginsection#1\par{\medbreak\bigskip
      \message{#1}\leftline{\bf#1}\nobreak\medskip
\vskip-\parskip
      \noindent}

\newcommand{\eq}{\begin{equation}}

\newcommand{\eqx}{\end{equation}}
\newcommand{\eqn}{\begin{eqnarray}}

\newcommand{\bi}{\begin{itemize}}

\newcommand{\eqnx}{\end{eqnarray}}
\newcommand{\ei}{\end{itemize}}

\newcommand{\nn}{\nonumber}
\newcommand{\ra}{\rangle}
\newcommand{\la}{\langle}

\newcommand{\kt}{\boldsymbol{k}}

\newcommand{\bt}{\boldsymbol{b}}
\newcommand{\xt}{\boldsymbol{x}}

\newcommand{\A}{\mathcal{A}}
\newcommand{\Hc}{\mathcal{H}}

\begin{document}
\begin{center}

\grande{Exploring an S-matrix for gravitational collapse}
\vskip 5mm
\vskip 5mm
\large{ G. Veneziano}
\vspace{3mm}

{\sl Theory Division, CERN, CH-1211 Geneva 23, Switzerland}

{\sl and}

{\sl Coll\`ege de France, 11 place M. Berthelot, 75005 Paris, France}

\vspace{6mm}

\large{ J. Wosiek}
\vspace{3mm}

   {\sl M. Smoluchowski Institute of Physics, Jagellonian University}

{\sl Reymonta 4, 30-059 Cracow, Poland}\\

 \vspace{6mm}
\centerline{\medio  Abstract}
\vskip 5mm
\end{center}
 We analyze further a recently proposed  S-matrix description of transplanckian scattering in the specific  case of axisymmetric  collisions of extended sources, where some of the original approximations are not necessary.  We confirm  the claim that such an approximate description appears to capture the essential features of (the quantum counterpart of) classical gravitational collapse. More specifically,  the S-matrix develops singularities whose location in  the sources'  parameter space  are consistent with (and numerically close to) the bounds coming from closed-trapped-surface collapse criteria.  In the vicinity of the critical ``lines" the phase  of the elastic S-matrix exhibits a universal fractional-power behaviour reminiscent of Choptuik's scaling near critical collapse.

 \vspace{5mm}

\vfill
\begin{flushleft}
CERN-PH-TH/2008-076 \\
TPJU - 3/2008\\
April 2008\\
\end{flushleft}
\vfill
\section{Introduction}
In a recent paper \cite{ACV07} (hereafter referred to as ACV) a
simplified non-perturbative treatment of transplanckian scattering
has been proposed. It is based on an approximate resummation of the
semi-classical corrections to the leading eikonal approximation.
These  were identified long ago \cite{ACV} as coming from  tree
diagrams  where the fast initial particles act as external  sources
for gravitons.

 Resumming tree diagrams amounts to solving the equations of motion of an effective action proposed
 quite sometime ago \cite{Lip, ACV93}. Taking the high-energy limit simplifies considerably
 the longitudinal dynamics which, in a certain approximation (neglect of so-called ``rescattering diagrams"),
 can be factored out leaving behind an effective dynamics in the ``transverse"
$(D-2)$-dimensional space. In \cite{ACV07} (and here) the case of $D=4$,
hence of an effective $two$-dimensional theory, was (will be) considered.

 In \cite{ACV07} one further simplification consisted of neglecting a physical polarization of
 the radiated gravitons, the one suffering from infrared problems. Finally, some azimuthal averaging was
 used in order to be able to treat the problem both analytically and numerically.
In spite of  these approximations, the resulting S-matrix looked
able to capture the essential physics  of the gravitational collapse
problem.  In particular, it showed the existence of a critical impact parameter $b_c$ below which
 a new absorption (and thus the opening of some new
channels) occurs. That critical value, possibly signalling a
transition between what we may call a dispersive (D) and a
black-hole (BH) phase,  turned out to be in good agreement with the
one based on closed-trapped-surface (CTS) criteria \cite{EG},
\cite{KV}.  A more recent investigation \cite{MO} managed to  solve
numerically the full partial differential  equations (PDE) (i.e.
without  using the azimuthal average approximation) and largely
confirmed   the  qualitative outcome  of the simplified  ACV
approach.

In this paper we consider more systematically the case of axisymmetric extended colliding sources/beams. The motivations for considering this particular case are several: on the one hand the PDE's reduce to ODE's making the problem an affordable one by analytic (or by much easier numerical) techniques, without having to make the azimuthal averaging approximation. At the same time, as we will show, the IR-sensitive polarization is simply not produced in the axisymmetic case. Last but not least,  we are able to introduce a vaste number of initial states, by playing at will with the many (shape and intensity) parameters chracterizing the sources and check for the existence of critical surfaces in this multidimensional space. The results can then be compared with those coming from CTS criteria \cite{KV} and, hopefully in the near future, will be tested against analytic \cite{analyt} as well as  numerical GR calculations \cite{numerical}.

The  paper is organized as follows: in section 2 we make some
general considerations on the axisymmetric case and prove a one-way
relation between the CTS criterion of \cite{KV} and the criticality
condition in the ACV system. In section 3 we give some examples of
interesting extended sources while in section 4 we determine
numerically the critical lines in the sources' parameter space. We
also compare those  lines with the bounds coming from the (much
simpler) CTS conditions and find very good agreement.  In section 5
we study the behaviour of the elastic S-matrix
in the vicinity of the critical points and point out a possible
connection with Choptuik's scaling \cite{Chop} near critical
collapse.
 Section 6 briefly summarizes our results and  future prospects.

\section{The axisymmetric case: general considerations}

Our starting point is the effective two-dimensional action of \cite{ACV07} (see their equation (2.22)):
\eqn\label{reduced_a}
\frac{\A}{2\pi Gs}~&=&~a(\bt)+\bar{a}(0)-\frac{1}{2}\int d^2\xt \nabla \bar{a}\nabla a+\frac{(\pi R)^2}{2}\int d^2\xt(-(\nabla^2\phi)^2+2\Hc \nabla^2\phi) \, ,  \nn \\
-\nabla^2 \Hc~&\equiv &~\nabla^2 a~\nabla^2\bar{a}-\nabla_i\nabla_j a~\nabla_i\nabla_j\bar{a}\, ,
\eqnx
where $a$, $\bar{a}$ and $\phi$ are three real fields representing the two longitudinal and the (IR-safe) transverse component of the gravitational field, respectively.
Equation (\ref{reduced_a}) can be easily generalized in order to deal with two extended sources:
\eqn
\label{Aexts}
\frac{\A}{2 \pi G s} &=& \int d^2x \left[a(x) {\bar s}(x) + {\bar a}(x) s(x) - \frac12 \nabla_i {\bar a} \nabla_i a \right] \nonumber \\
&+& \frac{(\pi R)^2}{2}  \int d^2x \left(- (\nabla^2 \phi)^2 + 2 \phi \nabla^2 {\cal H} \right)\, ,
\eqnx
where  the center of mass energy ${\sqrt s}$ provides the overall normalization factor $ 2 \pi G s = \frac{\pi}{2 G} R^2$, while the two sources $s(x), {\bar s} (x)$ are normalized by $\int d^2x ~s(x)= \int d^2x~ {\bar s} (x)  = 1$.

Let us now specialize to  the case of two extended axisymmetric sources moving in opposite direction with the speed of light and undergoing a central collision.
Using the conventions of \cite{ACV07} we will denote by $E_i(r_i)$ (in the following $i=1,2$ will represent unbarred and barred fields/sources respectively) the energy carried by the ith beam below $r = r_i$ and define $R_i(r) = 4 G E_i(r)$.
Let us also assume that the two sources have finite support so that $R_i(r) = R_i(\infty) \equiv R_i$
for $r > L_i$. By going to the overall center of mass, we may always choose $R_i = R = 2 G \sqrt{s}$.

\subsection{Simplifications}

One advantage of considering the axisymmetric case is that there is simply no dependence of the physics upon the azimuthal angle, hence no need to take averages over it. This is a useful technical simplification that allows us to reduce the problem to solving ODE.

The second more important advantage  comes from the  observation
that the IR-singular ``LT" graviton polarization is not produced in
that case. Thus the problem is completely IR-finite even in $D=4$.
In order to see this, let us recall from \cite{ACV07} that the LT
polarization is produced with an amplitude proportional to
$\sin\theta_{12} \cos\theta_{12}$. In the notations of \cite{ACV07}:
\eq A_{LT} = A_{\mu\nu} \epsilon^{\mu\nu}_{LT} \sim \kt^{-2} \sin
\theta_{12}\cos \theta_{12} \, , \eqx where $\theta_{12}$ is the
angle between the two transverse momenta $\kt_1, \kt_2$ that combine
to give a physical graviton of momentum $\kt$.  The angular factor
can be expressed in terms of the momenta as: \eq \kt_1^2\kt_2^2 \sin
\theta_{12}\cos \theta_{12} = \epsilon_{ij} \kt_1^i\kt_1^k \kt_2^j
\kt_2^k \, . \eqx Going over to position space this becomes a
differential operator acting on $a$ and $\bar{a}$: \eq
 \epsilon_{ij} \nabla^i \nabla^k a \nabla^j \nabla^k \bar{a} \, .
\eqx

Given that in the axisymmetric case $a$ and $\bar{a}$ depend only on $\xt^2$ it is easy to see that the above expression vanishes as a result of the antisymmetry of $\epsilon_{ij}$.

\subsection{ Axisymmetric action and equations of motion}

It is straightforward to rewrite the action (\ref{Aexts})  for the axisymmetric case as a one dimensional
integral over the variable $r^2 = x^2 \equiv t$. Using $\int d^2 x = \pi \int_0^{\infty} dt$ we find:
\eqn
\label{At}
\frac{\A}{2 \pi^2 G s} &=& \int dt \left[a(t) {\bar s}(t) + {\bar a}(t) s(t) - 2 \rho  \dot{\bar a} {\dot a} \right] \nonumber \\
&-& \frac{2}{(2\pi R)^2}  \int dt  (1- {\dot \rho} )^2\, ,
\eqnx
where a dot means $d/dt$ and, as in \cite{ACV07}, we have introduced the field:
\begin{equation}
\label{rhodef}
\rho = t \left( 1- (2 \pi R)^2 \dot{\phi} \right) \,.
\end{equation}
Integrating by parts and using  $\pi \int_0^t dt' s_i(t') = R_i(t)/R$ we arrive at the following convenient form of the action:
\eq
\label{Atconv}
\frac{\A}{\hbar} = - \frac{1}{4 \l_P^2} \int dt \left[  (1- {\dot \rho} )^2 - \frac{1}{ \rho} R_1(t) R_2(t)
+(2\pi R)^2 \rho \left({\dot a_1} +\frac{R_1(t)}{2 \pi R \rho}\right) \left({\dot a_2} +\frac{R_2(t)}{2 \pi R \rho}\right)   \right] \, .
\eqx

 The equations of motion that follow from (\ref{Atconv}) read:
 \eqn
 \label{aseqs}
 \dot{a}_i &=& - \frac{1}{2 \pi \rho}\frac{R_i(t)}{R} \, ,\nonumber \\
 \ddot{\rho} &=& \frac12 (2 \pi R)^2 \dot{a}_1 \dot{a}_2 = \frac12 \frac {R_1(t)R_2(t)}{\rho^2}\, ,
 \eqnx
 and  therefore reduce to a closed 2nd order equation for $\rho$. We want to look for solutions of that equation with the following boundary conditions \cite{ACV07}:
 \eq
 \rho(0) = 0 ~~,~~ \rho(t) \rightarrow t~~{\rm  as} ~~  t \rightarrow \infty \, .
 \eqx
 Given the finite support of the sources the latter condition can be replaced by the requirement:
 \eq
 \dot{\rho} = \sqrt{1-R^2/\rho} ~~~{\rm  for} ~~~ \sqrt{t} > Max(L_1, L_2). \label{asym} \, .
 \eqx

 For given source profiles $R_i(t)$ a possible strategy for solving the problem is to reduce it to a first order system:
 \eqn
 \label{firstorder}
 \dot{\rho} &=& \sqrt{\sigma - \frac{R_1(t)R_2(t)}{\rho}} ~~~  i.e. ~~ \sigma \equiv \dot{\rho}^2 + \frac{R_1(t)R_2(t)}{\rho} \, ,
 \nonumber \\
 \dot{\sigma} &=&\frac{1}{\rho} \frac{d (R_1 R_2) }{dt} \, ,
 \eqnx
with initial conditions
\eq
\rho(0) = 0~~, ~~  \sigma(0) = \sigma_0, \label{init} \, ,
\eqx
and to find a $\sigma_0$ such that $\sigma(Max(L_1,L_2)) = 1$.
For sufficiently small (large) $R_i/L_i$ one expects that the latter condition can (cannot) be imposed on real-valued solutions. Thus, in general, there should be some critical values for those parameters separating two different regimes.

\subsection{CTS-criteria and critical points: a general result}

In the general axisymmetric case, one can construct explicitly a minimal CTS \cite{KV} provided that an $r_c$ exists such that  (see eq. (4.4) of \cite{KV} for $D=4$):
\eq
\label{KVcrit}
R_1(r_c) R_2(r_c) = r_c^2\, .
\eqx

We will now argue that such a condition implies the absence of real solutions to eqns. (\ref{aseqs}) with $\rho(0) =0$.
Proof:
Let us first note that, because of (\ref{firstorder}) and the fact that the $R_i$ are non-decreasing functions of $r$, the quantity $\sigma$, as well as $\dot{\rho}$, are increasing functions of $t$. Therefore, for any $t$:
\eq
\label{ineq}
\sigma(t) \le \sigma(\infty) = 1 ~~, ~~ {\rm i.e.}~~ {\dot \rho}(t) \le \sqrt{1- \frac{R_1(t)R_2(t)}{\rho(t)}} \, .
\eqx
Assuming that the KV criterion (\ref{KVcrit}) can be met let us write:
\eq
\label{rho0bound}
\rho(0) = \rho(t_c) - \int_0^{t_c} dt' {\dot \rho}(t') > \rho(t_c) - t_c {\dot \rho}(t_c) >
\rho(t_c) - t_c  \sqrt{1- \frac{t_c}{\rho(t_c)}}\, ,
\eqx
where we have used eqs. (\ref{KVcrit}) and (\ref{ineq}). At this point it is easy to check that the rhs of
 (\ref{rho0bound}) cannot vanish for any  (positive) value of $\rho(t_c)$ thus proving that we cannot impose  the condition $\rho(0)=0$ when the criterion (\ref{KVcrit}) is satisfied.

\section{Examples of  source profiles}
\subsection {Two identical  two-parameter sources}

A  class of finite-size  sources is characterized by the following profiles:
\eq\label{acv}
s_1(t) = s_2(t) = \frac{d}{\pi \left(d + (1-d)t^2\right)^{3/2}}\Theta(1-t)\, ,
\eqx
where, without lack of generality, we have fixed the sizes of the two  beams to be 1.
One can easily verify that these sources satisfy :
\eq
\pi \int_0^t dt' s(t') = R(t)/R =  \frac{t}{ \left(d + (1-d)t^2\right)^{1/2}}   ~~, ~~ \pi \int_0^1 s(t)dt = 1\, .
\eqx

Inserting these expressions in our differential equation (\ref{firstorder}) leads to the equations:
 \eqn
 \label{firstorder1}
 \dot{\rho} &=& \sqrt{\sigma - \frac{R^2(t)}{\rho}} ~~~, ~~~ \dot{\sigma} = \frac{1}{\rho}\frac{d R^2}{dt}\, ,
  \nonumber \\
  R^2(t) &=& \frac{R^2 t^2}{d + (1-d)t^2} ~~, ~~ t<1\, ,
  \eqnx
while for $t >1$ we should impose $\sigma =1$.

This system can be easily studied numerically. In particular one can find how the critical value of $R$,
$R_c$,  depends on the parameter $d$. This parameter gives (for fixed total extent $L \equiv 1$) the ``shape" of the extended sources:
$d=1$ corresponds to the case of constant-density sources considered in \cite{ACV07} ,  $d <1$  to sources roughly concentrated around $r \sim \sqrt{d} L$, and $d>1$ to sources peaked around $r=L=1$.

\subsection {Central scattering of a particle off a ring}

In this, very asymmetric  case $R_1 = R$ while $R_2 = R \Theta(t-b^2)$. The notation anticipates  that we can view this case as an approximation to the scattering of two particles at impact parameter $b$. And indeed, from (\ref{aseqs}), we recover in this case the approximation used in \cite{ACV07} to describe the latter process i.e.
\eq
\label{ponring}
\ddot{\rho} = \frac12 \frac {R_1(t)R_2(t)}{\rho^2} = \frac12 \frac {R^2}{\rho^2}  \Theta(t-b^2)\, .
\eqx
If we require $\rho(0)=0$ this equation leads to the condition on $\rho(b^2)$:
\eq
\rho(b^2) = b^2 \dot{\rho}(b^2) = b^2 \sqrt{1- \frac{R^2}{\rho(b)}}\, ,
\eqx
which has a real solution only if $\frac{R}{b} < (\frac{R}{b})_c = 2^{1/2} 3^{-3/4} \sim 0.62$.
Such a result has to be compared with the upper limit given by (\ref{KVcrit}) which, in this case, simply becomes $(R/b)_c^{CTS} < 1$.

It is interesting to notice that exactly the same  $(\frac{R}{b})_c$ will apply to the situation in which the point-like source is replaced by an arbitrary source ``contained" inside the ring-shaped one. Physically this makes sense since, by Gauss'  theorem, the compact source  should propagate undisturbed while the more extended source is only affected by the total energy of the more compact one.

\subsection {Gaussian sources}
Point-like sources are difficult to deal with numerically (especially in  momentum space). Therefore
we also introduce Gaussian-smeared versions of the above point and ring-like  sources.
\eqn
s_1(\xt)=\frac{1}{{\cal N}_1} \exp{\left(-\frac{r^2}{2\sigma^2}\right)}\Theta&(&L_1-r) ~,~ s_2(\xt)=\frac{1}{{\cal N}_2}
\exp{\left(-\frac{(r-L_2)^2}{2\sigma^2}\right)}\Theta(L_2-r),\;\;\; \nonumber \\
{\cal N}_1=2\pi\sigma^2(1-\exp{\left(\frac{-L_1^2}{2\sigma^2}\right)}) &,&
{\cal N}_2=2\pi\left(\sigma^2(\exp{\left(\frac{-L_2^2}{2\sigma^2}\right)}-1)+
\sigma L_2 \sqrt{\frac{\pi}{2}}   \rm{Erf}{\frac{L_2}{\sqrt{2}\sigma}}
\right)\; .
\eqnx
When $\sigma\longrightarrow 0 ~ (\infty)$ the problem  reduces to the one of the point-ring (two homogeneous beams) case.

Another interesting example is the one of gaussian sources concentrated at $r=0$. They correspond to:
\eq \label{gso}
s_i(t) = \frac{1}{2\pi L_i^2} exp \left(-\frac{t}{2L_i^2}\right)~~,~~ \frac{R_i(t)}{R} = 1- exp \left(-\frac{t}{2L_i^2}\right)\, .
\eqx

\section{Numerical solutions and  critical lines}

Solving Eqs.(\ref{aseqs}) serves  two purposes: a) to determine a critical line which separates the dispersive (D) and black-hole
(BH) phases, and b) to obtain the actual solution, for a given source parameters, and to calculate various on shell observables.
Let us illustrate the first procedure in the case of the scattering of the two identical beams with
the sources
(\ref{acv}). There are two independent parameters: $d$ - specifies the shape of the energy distribution, and $R/L$ the ratio of
the Schwarzschild radius of a source to its maximal extent $L$.  We shall work in units of $L$.
As already outlined in Sect.2.1 we would like to find an initial value $\sigma_0$, in (\ref{init}), such that the corresponding
 solution satisfies $\sigma(1)=1$, matching the $t>L^2$ behaviour (\ref{asym}). We shall refer to such a solution as the maximal solution.

\begin{figure}[h]
\epsfig{width=10cm,file=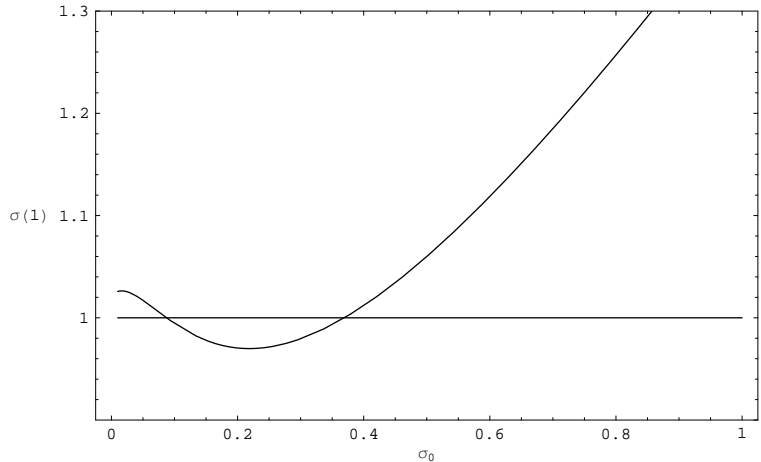} \vskip-4mm \caption{Looking for
the maximal solution of Eqs.(\ref{firstorder}), $\sigma(1) = 1$, as
a function of the initial value $\sigma(0)=\sigma_0$ (in the
specific case $d=1$, $R/L=0.46$) }
  \label{g1ofg0}
\end{figure}

To this end, we examine the dependence $\sigma(1)$ on $\sigma_0$, at given $(d,R)$, cf. Figure \ref{g1ofg0},
and identify the region, in $(d,R)$, where the equation $\sigma(1,\sigma_0)=1$ has a real solution.
It turns out that when we increase $R$, at fixed $d$, the minimum in Fig.\ref{g1ofg0}, moves upwards giving rise to  three
cases: i) for $R < R_1$ there is only one solution, ii) for $R_1 < R < R_c$ there are  two (real)  solutions, and iii) for $R_c < R$ there are no real solutions.
We then determine the critical values $R_c$ for a range of $d$ thus obtaining the critical
line $R_c(d)\equiv(R/L)_c(d)$ in the
$(d,R/L)$ space. All numerics is done with the aid of Mathematica.

\begin{figure}[h]
\epsfig{width=10cm,file=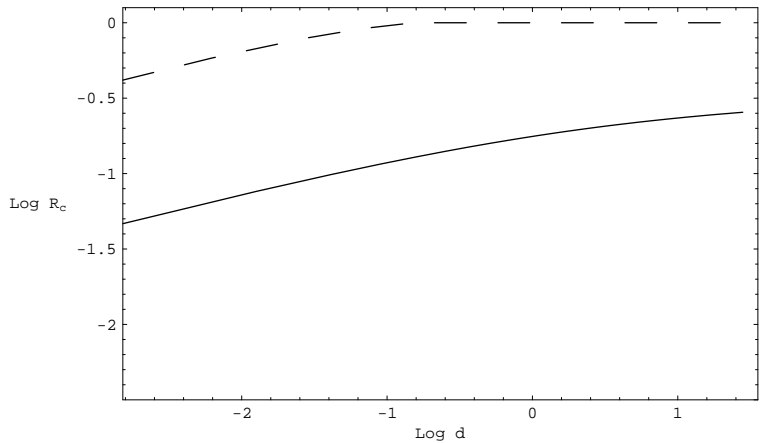} \vskip-4mm \caption{The
critical (solid) line  in the $(R,d)$ plane having set $L=1$. We
also show (dashed line)
 the upper bound on $R_c$ from the CTS criterion
(\ref{KVcrit}). The  BH phase is above the solid line.}
  \label{Rcofd}
\end{figure}

The final result, cf. Fig.\ref{Rcofd},
confirms well our expectations. In the limit of the homogeneous sources $( d=1 )$ we recover $R_c=0.47$ -- the value already quoted  in \cite{ACV07}. For $ d \longrightarrow \infty $ one expects to reach the kinematics of the particle-ring scattering, with
$R_c=0.62$, and indeed the critical line is consistent with this prediction.
In  Fig.\ref{Rcofd} we also plot the upper bound on $R_c$ coming from the CTS criterion (\ref{KVcrit}).
It is easy to check that this reads:
\eqn
\label{CTSd}
R_c^{CTS} =  \left(4 d(1-d)\right)^{1/4} \Theta (1/2-d) +  \Theta (d- 1/2)\, .
\eqnx

\vspace*{0.5cm}

We have also checked that, to a good approximation,  our  $R_c$ roughly varies as $d^{1/4}$ for
small $d$ in nice (though not necessary) agreement with (\ref{CTSd}).

As a second example  consider the scattering of two Gaussian sources, concentrated at $\xt=0$ with
transverse sizes $L_1, L_2$,  as in (\ref{gso}). Strictly speaking, the sources do not have a finite support,
however they vanish quickly for $t  \gg  L_i $. Therefore we replace the condition $ \sigma(1)=1 $ with
$\sigma( \# Max(L_1,L_2))=1 $. In practice taking $ \# = 10 $ is more than sufficient.
This time we set $R=1$ and examine the transition in the $(L_1, L_2)$ plane. The result is shown in
 Fig.3. Obviously, the critical line is symmetric with respect to $ L_1 \leftrightarrow L_2 $.
 It's shape corresponds roughly to a straight line $L_1 + L_2 = const$
 (as clearly visible in a linear plot not shown here), suggesting
 that this variable controls
the effective concentration of the total energy. Once more we can ask how this critical line compares
with the lower bound one would find using the CTS criterion
(\ref{KVcrit}).
This can be easily done numerically and the result is  shown again in Fig.3.
 Amusingly, also the CTS criterion gives a curve which, in the central region $L_1\sim L_2$, roughly
 corresponds to $L_1 + L_2 = const$.
 Besides a typical factor 2 discrepancy (but remember: CTS criteria only give bounds!), the two curves start to diverge when the two $L_i$ are very different. In fact, while the CTS-curve goes to the point $L_1=0, L_2 = \frac{1}{\sqrt2}$, our result indicates that this point moves to infinity as $L_1 \rightarrow 0$,  i.e. that one is always in the BH regime in that limit.

\begin{figure}[h]
\begin{center}
\epsfig{width=10cm,file=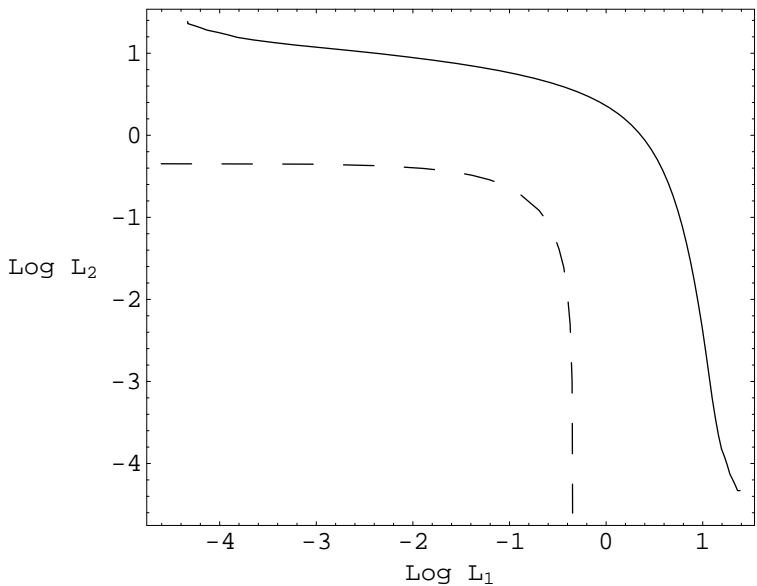} \vskip-4mm \caption{Two gaussian
sources around the origin. The BH phase lies below the critical (solid) line  in the $(L_1,L_2)$
plane. We also show (dashed line)
the lower bound on the curve  from the CTS criterion
(\ref{KVcrit}).}
\end{center}
\label{L2of}
\end{figure}

\begin{figure}[h]
\begin{center}
\epsfig{width=10cm,file=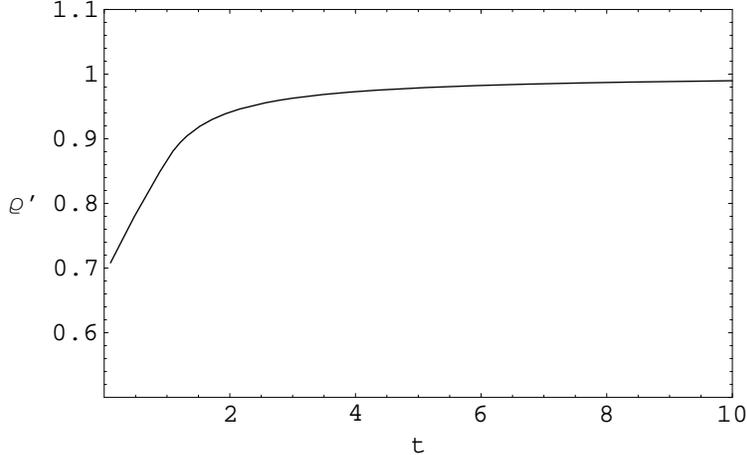}
\vskip-4mm \caption{The maximal solution, in the dispersive phase,
also for $t>1$ (two identical sources, eq. (\ref{acv}),  with $d=1$, $R/L=0.44$).}
\end{center}
\label{rhopoft}
\end{figure}

All the above calculations required solving Eqs.(\ref{firstorder}) for $t \le 1$. Next we look for the maximal solution in the
larger domain $ t \le t_{max} $. To this end it suffices to use the value $\sigma_0$ found above and extend the
profile $R_i(t)$ beyond $L^2$: $R_i(t)=R, t> L^2$. The rest is again done by Mathematica. As an example, we show
in Fig.\ref{rhopoft} the derivative of the solution $\rho(t)$. It interpolates smoothly around $t=1$ and, as expected,
tends to 1 at large $t$. We have also verified that for $t > 1$,
$\rho(t)$ obtained above is identical to the analytic solution for the constant profiles
\eqn
\rho_{an}(t)= R^2 F^{-1}\left[ F(\rho_0/R^2) + t/R^2 - t_0/R^2\right],
\eqnx
with $\rho_0=\rho(t_0),\;\;\; t_0>1$, and
\eqn
F(x)= \sqrt{x(x-1)}+log(\sqrt{x}+\sqrt{x-1}).
\eqnx

\section{Behaviour  near the critical point}
An interesting quantity to compute is obviously the on-shell action, i.e.  the value of the action on the e.o.m. as a function of the external parameters (the profiles of the external sources).
In the multidimensional parameter space we expect to find, as in the collapse problem, a critical surface of co-dimension 1 where the on-shell action itself develops some branch-point singularity.
Typically, on the other side of that critical surface the action acquires an additional imaginary part (on top of the one due to graviton emission) providing  an additional absorption of the elastic amplitude \cite{ACV07}.

This extra absorption could possibly be interpreted as the opening-up of new channels e.g. those corresponding to the formation of black holes behaving as stable particles at leading order in $\hbar$.
The absorption itself is expected to be related (up to some numerical factor) to the number of new channels that are opening up beyond criticality and to provide therefore information on the entropy of the black holes that are supposedly formed. Under these assumptions a connection can possibly be established between the critical exponents of the action and those of Choptuik's scaling \cite{Chop}.

On the equations of motion the   action (\ref{Atconv}) immediately simplifies to read:
\eq
\label{Atconveom}
\frac{\A}{\hbar}^{eom} = - \frac{1}{4 \l_P^2} \int dt \left[  (1- {\dot \rho} )^2 - \frac{1}{ \rho} R_1(t) R_2(t)
  \right] \, ,
\eqx
while using the second of eq (\ref{aseqs}) the above expression can also be written in the form:
\eqn
\label{Axeom}
 \frac{\A}{\hbar}^{eom} &= &- \frac{1}{4 \l_P^2} \int dt \left[  (1- {\dot \rho} )^2 - 2 \rho {\ddot \rho} \right]
= - \frac{1}{4 \l_P^2} \int dt \left[3 (1- {\dot \rho} )^2 - 2 t {\ddot \rho} -2 \frac{d}{dt}[(t- \rho)(1- {\dot \rho} )]\right] \nn \\
 &=&
- \frac{1}{4 \l_P^2} \int dt \left[3 (1- {\dot \rho} )^2 - 2 t {\ddot \rho} \right]\, .
\eqnx
Here we have used the fact that the last surface term vanishes because of the properties of the solution at $t=0$ and $t= \infty$.

We would like to study the action near the critical point where we
expect to find a branch point behaviour. Unfortunately, the action itself is IR-singular:
the infinite Coulomb phase, even if  it is unobservable, has to be subtracted carefully in a numerical
analysis in order to avoid that interesting finite terms are subtracted as well.
Also, the branch point appears to be quite soft: a power $(p-p*)^{3/2}$ ($p$ being a generic source parameter with $p*$ its critical value)
behaviour which has to be
unraveled from leading analytic pieces (a constant and a linear term).

Fortunately, a trick can be found in order to avoid the IR-singularity. Consider the partial derivative
of the
on-shell action wrt a particular parameter of the problem, e.g. the total cm energy. In general, this derivative will receive two contributions: one from the {\it explicit} dependence of the action from that parameter; and one from the {\it implicit} dependence of the solution (the fields) upon the parameter. However, this second contribution is also proportional to the variation of the action with respect to the fields which, by definition of the classical equations, vanishes on shell. We are thus led to:
\eq
\frac{\partial {\A}}{\partial p} = \int \left(\frac{\partial L}{\partial p}\right)_{\rm{fixed~ fields}}\, .
\eqx

We can use this general strategy to our advantage in the case at hand by considering derivatives of
$\frac{\A}{Gs}$ and, as parameter $p$,
the overall cm energy $R^2$ while keeping the shapes of the two sources, $R_i(t)/R$  fixed. Since:
\eq
 \frac{\A}{Gs} =  -  \int dt \left[  \frac{(1- {\dot \rho} )^2}{R^2} - \frac{R_1(t) R_2(t)}{R^2 \rho}
+(2\pi)^2 \rho \left({\dot a_1} +\frac{R_1(t)}{2 \pi R \rho}\right) \left({\dot a_2} +\frac{R_2(t)}{2 \pi R \rho}\right)   \right]\, ,
\eqx
we see that only the first term contributes to the abovementioned partial derivative so that:
\eq
\label{xAder}
\frac{\partial {(\A/Gs)}}{\partial R^2} = \frac{1}{R^4} \int dt  (1- {\dot \rho} )^2\, ,
\eqx
which is now perfectly IR-finite.

Interestingly, the above derivative of the action is directly related to the average multiplicity of
emitted gravitons. Using the effective form of the complex action introduced in \cite{ACV07} one readily obtains:
\eq
\frac{\partial
{(\A/Gs)}}{\partial R^2} = \frac{\pi^2}{R^3 \sqrt{s}} \la N \ra \, ,
\eqx
which shows that the total multiplicity is in fact IR-finite in the axisymmetric case (recall that the IR-sensitive polarization is not produced in this case).

\begin{figure}[h]
\begin{center}
\epsfig{width=11cm,file=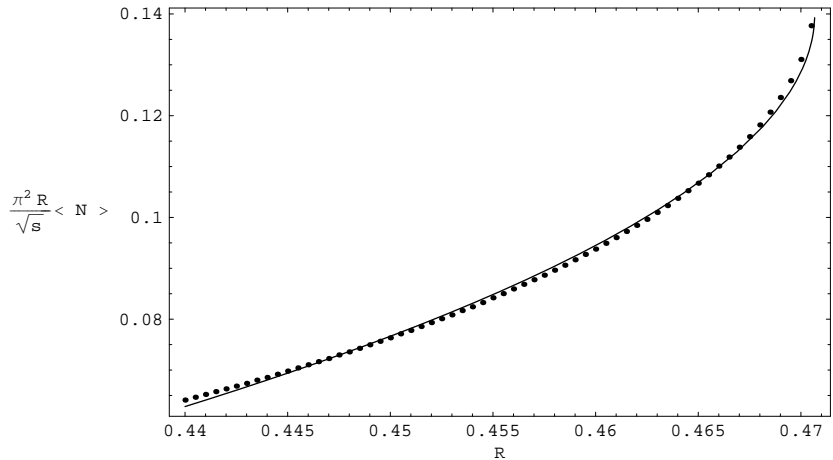}
\end{center}
\vskip-4mm \caption{The total multiplicity of emitted gravitons (points) and
the best fit: $0.138-0.46(R_c-R)^{0.523}$. A fit with the fixed
power $1/2$ is marginally worse.}
  \label{nav}
\end{figure}

We are now ready to analyze numerically the behaviour of the on
shell action in the vicinity of the critical point. In Fig. \ref{nav}
we show $R^4 \partial_{R^2} (\A/Gs) $, as obtained from the integral
(\ref{xAder}) in a range of $R$'s. At first sight one might suspect
that $\la N \ra$ is infinite exactly at $R_{c}$. A careful analysis
shows that it is instead finite although it approaches the critical point
with an infinite derivative. The fit, shown by a solid line, is
consistent with a square root branch point, \eqn \la  N \ra = c_0
+c_1 (R_c-R)^\frac{1}{2}\, , \eqnx in the vicinity of $R_c$. This
confirms the validity of the expansion \eqn \label{actionfit}
A(R)=A_0+A_1(R_c-R)^b+A_2(R_c-R)^c\, , \eqnx with $b=1$ and $c=3/2$.
Such a behaviour, first found in \cite{ACV07} and later confirmed in
\cite{MO}, could  be possibly related \cite{ACV07} to a
Choptuik-like \cite{Chop} dependence of the mass of the barely
formed black hole from $(R-R_c)$. In that case our result would be interpreted as meaning that  $M_{BH} \sim (R-R_c)^{\gamma}$,
with $\gamma = 3/4$, an exponent which is about twice the one found by Choptuik \cite{Chop} for the spherical collapse of some fluids. It will be very interesting to see what values of $\gamma$ will come out of future numerical studies of the collapse of extended sources of the type discussed here.

\begin{figure}[h]
\begin{center}
\epsfig{width=10cm,file=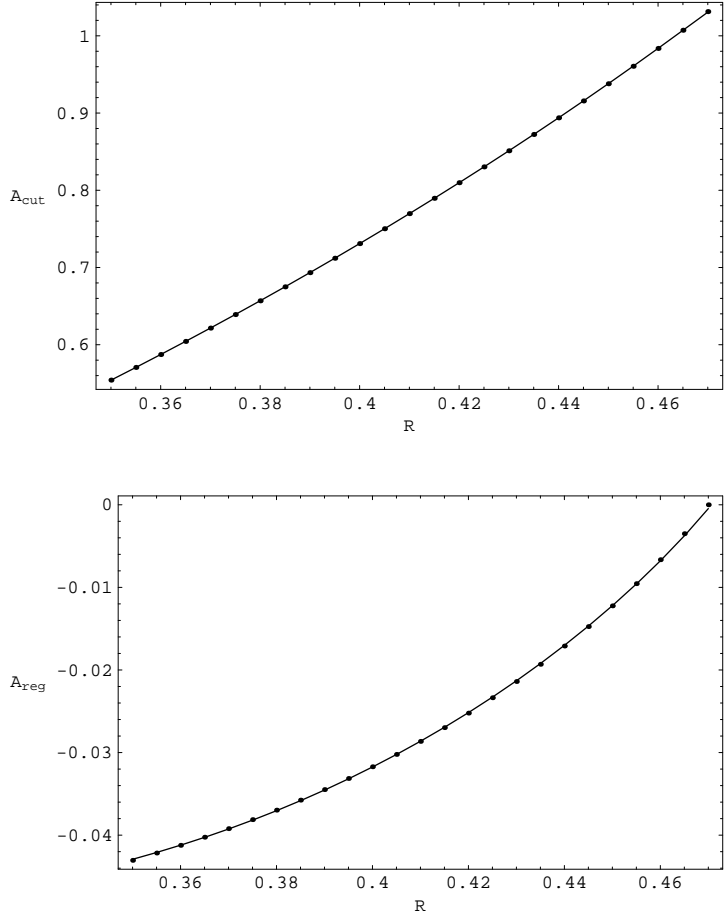}
\end{center}
\vskip-4mm \caption{Fitting the critical behaviour of the cutoff (upper, $t_{max}=50$) and the subtracted
actions. }
  \label{actionfits}
\end{figure}

One can also check directly the  behaviour of the full, suitably regularized, action (\ref{Atconveom})
close to the transition point. We have used two regularizations: 1) the integral was cutoff  at some large
value $t_{max}$, defining the cutoff  action $A_{cut}$, and 2) we have introduced a counterterm,
$1/\rho(t) \longrightarrow 1/\rho(t) - 1/t$, which subtracts the infrared divergence and results in the
IR and UV-finite expression $A_{sub}$. Figure \ref{actionfits} shows these actions together with  fits
like (\ref{actionfit}) and two {\em a priori} unconstrained  powers.

Comparing the two plots one sees immediately that the cutoff  action is much more
linear than the subtracted one. What happens is that $A_{cut}$ is obviously much bigger
(cf. the scales of both figures) and the
nonlinear effect of the last term is relatively weaker. On the other hand, in $A_{sub}$ the big "background"
is already subtracted exposing better the last two terms. This observation is also confirmed by our fits.
Simple Mathematica fitting routines were unable to disentangle two different powers in
$A_{cut}$, while fitting $A_{sub}$ gave $b=1.07\pm 0.12$ and $c=1.34\pm 0.22$ in good agreement with $b=1$
and $c=1.5$ predicted in \cite{ACV07}. As for $A_{cut}$, we were glad to find that a slightly less
general form,
\eqn\label{afit2}
A(R)=A_0+A_1(R_c-R)^b+A_2(R_c-R)^{b+1/2}\, ,
\eqnx
works well, cf. Fig.\ref{actionfits} (upper part).
Finally, we have also checked the cutoff dependence of these parameters, see Table \ref{afit}.
\begin{table}[h]
\begin{center}
\begin{tabular}{ccccc}
 \hline\hline
  $t_{max}$   &  $   b    $  & $  A_0  $ &  $ A_1/A_0   $ & $A_2/A_1$ \\
   \hline
  20          &  $  1.044  $  & $ 0.826 $ &  $ -6.149 $ &   $-0.894 $ \\
  50          &  $  1.045  $  & $ 1.034 $ &  $ -6.096 $ &   $-0.883 $   \\
 100          &  $  1.046  $  & $ 1.189 $ &  $ -6.077 $ &   $-0.877 $     \\
 300          &  $  1.049  $  & $ 1.433 $ &  $ -6.060 $ &   $-0.871 $     \\
   \hline\hline
\end{tabular}
\end{center}
\caption{Parameters of the fit (\ref{afit2}) for a range of cut-offs, $t_{max}$}
\label{afit}
\end{table}

 As expected the power $b$
is cutoff independent and close to the predicted value $b=1$. The amplitudes $A_i$ show a mild, possibly logarithmic, dependence on $t_{max}$.
Interestingly, the ratios of these amplitudes seem to depend very weakly on $t_{max}$ suggesting a possible factorization
of the IR divergent contributions.

\section{Summary and future prospects}
In this paper we have continued the ambitious program initiated in ACV \cite{ACV07} by focussing on the  case of axisymmetric collisions. Both the ACV approach and the collapse criteria based on the explicit construction of minimal CTS \cite{EG,KV} greatly simplify  in this case. In the former the axial symmetry removes the infrared problem and reduces the equations to ordinary differential equations, often allowing for  quasi-analytical solutions. The CTS criteria also take a much simpler form, as exemplified by eq. (\ref{KVcrit}).

In spite of its simplicity, this case becomes extremely rich if one generalizes the problem of particle-particle scattering to the one of the central  collisions of two extended sources. The latter problem depends, in principle, on an infinite number of parameters (describing the energy profile of each beam) and thus allows  to study, in principle, how the information about the initial state is transmitted to the final state.

In this first paper we have concentrated our attention
on the relevant system of equations in position space, on the determination of the critical points/surfaces, and on the comparison with CTS-based results.
Our results have confirmed a rather striking connection between CTS collapse criteria  and the existence of critical points (in general hypersurfaces) in the parameter space describing the two incoming sources.
Not only could we  prove a general theorem that the CTS criterion of \cite{KV} implies the absence of real regular solutions of the ACV equations; we have also found that the bounds on the critical values based on \cite{KV} are typically only a factor two off the ACV predictions.

We have also investigated the behaviour of the on-shell action, hence of the S-matrix, as one approaches the critical points, confirming the robustness of a definite fractional power-law behaviour for the phase of the S-matrix. Within some assumptions this might be interpreted as a critical "Choptuik exponent" \cite{Chop} for the behaviour of the mass of the  black hole as one approaches the critical point from the collapse side.

Unfortunately, techniques for going beyond the critical points are still rudimentary, while the interpretation of the extra absorption of the elastic amplitude as being due to the opening of new (black-hole formation?) channels remains very speculative.
A full understanding of the role of inelasticity (i.e. of graviton emission) even in the subcritical regime is still necessary and might shed light of the nature of the "phase transition"  between a "full dispersion" regime and one in which a fraction of the incoming energy collapses.
For all these questions a momentum-space approach (see again \cite{ACV07}) --rather than the one in position space adopted here-- may turn out to be more suitable. We plan to present  it in a near-future investigation.

\section*{Acknowledgements}
GV wishes to acknowledge enlightening discussions with D. Amati and
M. Ciafaloni. JW is grateful to P. Bizon for instructive
discussions. We both enjoyed very useful exchanges with G.
Marchesini and E. Onofri and wish to thank them for communicating their results to
us  prior to publication.

\end{document}